\documentclass[aps,prl,preprint,nopacs,superscriptaddress]{revtex4-1}
\usepackage{graphicx}
\usepackage{verbatim}
\usepackage{mathrsfs}
\usepackage{amsmath,amsfonts,amssymb}
\usepackage{graphicx}
\usepackage{float}

\def\3{2.8in}    
\def\2{2.5in}
\def\4{3.0in}

\def \beq {\begin{equation}}
\def \eeq {\end{equation}}
\begin{document}

\title{Quasiparticle interference on type-I and type-II Weyl semimetal surfaces: a review}

\author{Hao~Zheng\footnote{haozheng1@sjtu.edu.cn}}
\affiliation {School of Physics and Astronomy, Shanghai Jiao Tong University, Shanghai, 200240, China}
\affiliation {Department of Physics, Princeton University, Princeton, New Jersey 08544, USA}

\author{M. Zahid~Hasan\footnote{mzhasan@Princeton.edu}}
\affiliation {Department of Physics, Princeton University, Princeton, New Jersey 08544, USA}



\begin{abstract}
Weyl semimetals are a new member of the topological materials family, featuring a pair of singly degenerate Weyl cones with linear dispersion around the nodes in the bulk, and Fermi arcs on the surface. Depending on whether the system conserves or violates Lorentz symmetry, Weyl semimetals can be categorized into two classes. Photoemission spectroscopy measurements have confirmed the TaAs-class and WTe$_2$-class of Weyl semimetals as type-I and type-II respectively. This review article aims to elucidate and elaborate on the basic concepts of Weyl semimetals and quasiparticle interference experiments on both type-I and type-II Weyl semimetals. The versatile results which reveal 1) the topological sink effect of the surface carriers, the unique feature of the Fermi arc state; 2) the weakly bound nature of the Fermi arc surface state; 3) the orbital dependent scattering channels on the surface; 4) a mirror symmetry protected surface Dirac cone, are summarily discussed. Finally, a perspective toward the future applications of quasiparticle interference techniques on topological materials is presented.

\end{abstract}
\date{\today}
\maketitle

\section{Introduction}
In the 1980s, the discovery of the integer and fractional quantum Hall effects, which arise from high mobility two-dimensional electron gases under high magnetic field, opened a new era of condensed matter physics \cite{QH1, QH2}. The previously successful Landau theory of phase transitions failed to describe these quantum Hall systems, as no symmetry is spontaneously broken. It was later recognized that the nontrivial topological Chern number induced by the Landau quantization of the Block wavefunction could explain the edge conductance in a quantum Hall system \cite{QH3}. Since the discovery of the quantum Hall system as the first nontrivial topological phase, the search for nontrivial topological phases arising from the intrinsic band structure of a crystal without an external high magnetic field has become a vital task for physicists. More than two decades later, the quantum spin Hall effect was experimentally discovered in strong spin-orbit-coupling (SOC) HgTe/(Hg, Cd)Te heterostructures \cite{QSH}. Theoretical research has revealed that the $Z_2$ topological index in a two-dimensional quantum spin Hall system can be generalized to three-dimensions, leading to the emergence of the first bulk topologically non trivial phase, the topological insulator (TI) \cite{TIthoery1, TIthoery2, TIthoery3}. The topology in a TI manifests itself in its bulk-boundary correspondence. More precisely, the bulk electronic band in a three-dimensional TI is gapped, while the surface is metallic. The TI phase was first discovered in the strong SOC material Bi$_{1-x}$Sb$_x$ alloy \cite{BiSbFu, BiSbHasan1, BiSbHasan2}, and the Bi$_2$Se$_3$-class of materials, which consist of Bi$_2$Se$_3$, Bi$_2$Te$_3$, and Sb$_2$Te$_3$ \cite{BiTeHasan1, BiTeZhang, BiTeHasan2, BiTeHasan3, BiTeChen}. The research potential of the latter was immediately realized, as its surfaces possesses only a single spin-momentum locked two-dimensional Dirac cone type electronic band, which can be treated as an unconventional type of two-dimensional electron gas which had not yet been discovered in any real two-dimensional material. Based on this novel surface state, many important effects, e.g. weak anti-localization \cite{WAL}, Landau quantization \cite{LLXue,LLHanaguri}, half-integer quantum Hall effect \cite{TIQH}, and the generation of Floquet-Bloch states under external excitations \cite{TIGedik} have been experimentally explored.  Furthermore, by incorporating magnetic order into a TI, it is possible to discern the mass acquisition of the surface Dirac fermions, as well as their hedgehog spin texture \cite{MTI1,MTI2}. A most remarkable achievement built upon the physics of magnetic TIs is the realization of the quantum anomalous Hall effect \cite{QAH1, QAH2, QAH3}, a milestone of condensed matter physics. Another fascinating research direction is superconducting TIs. By introducing a $s$-wave pairing to the surface state of a TI, a two-dimensional topological superconductor with unconventional $p_x+ip_y$ superconducting gap symmetry is effectively realized \cite{TSFu, TSJia1, TSXu, TSJia2}. At the boundaries where time reversal symmetry is broken, Majorana-type bound states have been successfully identified \cite{MF1, MF2}, paving the way toward the realization of non-Abelian anyons and topological quantum computation. In an inversion symmetric TI, the topology is characterized by a $Z_2$ index, which is determined by the number of band inversions at Kramers points and the SOC induced gap in the electronic band. Topological Kondo and Anderson insulators can be categorized in the same class as TIs as they share the same topological invariant $Z_2$ index \cite{TKI, TAI, TAI2}. However, their difference lies in the gap opening mechanism, where the Kondo effect and Anderson localization play the role of SOC in TIs.

The second symmetry protected topological phase in three dimensions is the topological crystalline insulator (TCI) \cite{TCI}, which has been discovered in the SnTe-class of materials \cite{SnSe, SnSeXu, SnSeStory, SnSeAndo}. Unlike the time reversal symmetry protected phase in TIs, the TCI is protected by mirror symmetry and is characterized by a mirror Chern number topological invariant. Therefore, a lattice distortion can add a mass term into the surface Dirac cone \cite{TCImass} and may induce a pseudo-magnetic field to the surface state \cite{TCIfield}. Theoretical calculations have predicted a novel large Chern number quantum anomalous Hall effect on magnetic TCIs \cite{TCIQH}.

The third discovered three-dimensional symmetry protected topological phase moves beyond the realm of insulators. The topological Dirac semimetal is a three-dimensional counterpart to graphene, which has been discovered to exist in Na$_3$Bi and Cd$_3$As$_2$ \cite{TDSMRappe, TDSMWang1, TDSMWang2, TDSMNagaosa, NaBiChen, NaBiXu, CdAsHasan, CdAsChen, CdAsCava}. The gapless nature of the bulk band of Dirac semimetals brings rich physics into both the two- and three-dimensional electron gases. Indeed, the giant magnetoresistance and chiral anomaly effect were observed in the bulk electron band \cite{TDSMCA1, TDSMCA2}, while unconventional quantum oscillations were detected in the topological surface state \cite{FermiArcQOTheory, FermiArcQOExperiment}. In addition, tip and pressure induced superconductivity in a topological Dirac semimetal has been shown, shedding a light on the realization of a new type of topological superconductor \cite{TDSMSC1, TDSMSC2, TDSMSC3}. Nanostructures of Dirac semimetals have also revealed significant Aharonov-Bohm oscillations \cite{TDSMAB}.

Beyond the search for new types of symmetry protected topological phases, the study of Weyl semimetals is strongly driven by the search for the long sought after Weyl fermions. In 1929, H. Weyl found that a massless Dirac fermion can be decomposed into a pair of relativistic particles with opposite chirality; the Weyl fermion. To date, Weyl fermions have yet to be discovered as fundamental particles in high energy physics. Recently, condensed matter physicists have discovered that in certain crystals that lack either space inversion or time reversal symmetry, their low-energy quasiparticle excitations can be described by the Weyl equation \cite{WeylRevHasan1, WeylRevYan, WeylRevHasan2, WeylRevBurkov}. Such a crystal has been termed as a Weyl semimetal. The bulk band structure in a Weyl semimetal features pairs of singly degenerate linearly dispersed Weyl cones of opposite chiralities. This chirality can be viewed as the fourth inherent property of a quasiparticle hosted in a crystal other than the charge, spin, and valley degree of freedoms. This leads to many unique Weyl fermion related transport effects such as the Adler-Bell-Jackiw anomaly \cite{WeylCATheory1, WeylCATheory2, WeylCATheory3}, the axial anomaly \cite{WeylAATheroy}, non-local transport \cite{WeylNT1, WeylNT2}, and the chirality dependent Hall effect \cite{WeylQH}. These novel phenomena uncover the rich correspondence between high energy particle physics and low energy condensed matter physics. In addition, the nontrivial topology in a Weyl semimetal can be characterized by the value of chiral charge, distinct from the topological invariants in TIs ($Z_2$ index) and in TCIs (mirror Chern number). Thus, the Weyl semimetal is identified as a new class of symmetry protected topological phase. The bulk-boundary correspondence in a Weyl semimetal manifests itself as an open contour on the surface, the Fermi arc surface state. Unconventional quantum oscillations induced by these Fermi arcs have been predicted \cite{FermiArcQOTheory} which serve as a key transport feature of this new type of surface state. Early attempts in the search for a Weyl semimetal in real materials were focused on time reversal symmetry breaking crystals Y$_2$Ir$_2$O$_7$ \cite{WeylWan} and  HgCr$_2$Se$_4$ \cite{WeylFang}, TI/trivial insulator multiple-layers \cite{WeylBalents}, as well as the solid alloy TlBi(S$_{1 - x}$Se$_{x}$)$_2$ \cite{WeylLin} and Hg$_{1 - x -y}$Cd$_x$Mn$_y$Te  \cite{WeylQi}. Unfortunately, unusual spin texture, complicated magnetic domain structure, difficulty in preparation, and the extremely strict requirements in fine tuning the stoichiometry in these material systems rendered the realization of these predictions fruitless.

The breakthrough in the search for a Weyl semimetal emerged from the TaAs-class of materials \cite{TaAsTheoryLin1, TaAsTheoryDai, TaAsTheoryLin2, TaAsTheoryYan}. The four members of this family share the same space group $I4_1md$ and a similar electronic band structure which features 12 pairs of Weyl nodes in the bulk bands and multiple Fermi arcs on the surface. Angle resolved photoemission spectroscopy (ARPES) measurements have explicitly proved TaAs \cite{TaAsARPESHasan1, TaAsARPESDing1, TaAsARPESChen1, TaAsARPESDing2}, NbAs \cite{TaAsARPESHasan2, TaAsARPESChen2}, TaP \cite{TaPARPESHasan, TaPARPESShi}, and NbP \cite{NbPARPESFeng, NbPARPESHasan, NbPARPESAndo, NbPARPESShi} as Weyl semimetals by direct observation of the linear dispersion of bulk Weyl cones and the arc shaped surface states which terminate at the projected Weyl points on the surface. Several key chiral Weyl fermion induced phenomena have also been observed in electronic transport and optics measurements, such as the extremely large magnetoresistance \cite{NbPTransportYan}, the helicity-protected ultrahigh mobility \cite{NbPTransportXu}, the violation of Ohm’s law \cite{NbPTransportKim}, the magnetic-tunneling-induced Weyl node annihilation \cite{TaPTransportJia}, the giant anisotropic nonlinear optical response \cite{TaAsOpticWu}, and the optical detection of Weyl fermion chirality \cite{TaAsOpticMa} have been experimentally discovered. Among these experiments, two exceptionally remarkable achievements lie in the detection of the signature of the chiral anomaly effect \cite{TaAsCAChen, TaAsCAJia} and the axial-gravitational anomaly effect \cite{NbPAA}, which may open a new era of table-top experimental realizations of high energy physics. Furthermore, the unusual spin texture of the Fermi arc surface state on TaAs has also been detected \cite{TaAsARPESSpinDing, TaAsARPESSpinHasan}.

Shortly after the discovery of the TaAs-class of materials, theory predicted the WTe$_2$-class of layered compounds to be type-II Weyl semimetals \cite{WTetheory, MoTetheoryChang, MoTetheoryYan, MoTetheoryBernevig}, and identifying the TaAs-class of Weyl semimetals to be type-I. Type-II Weyl fermions can be viewed as a tilted cone shaped dispersion in momentum space which breaks Lorentz symmetry and thus cannot exist as a fundamental particle in nature. This type of Weyl fermion exists as a unique phenomenon of condensed matter physics. Distinct from type-I Weyl semimetals, many novel effects such as the intrinsic anomalous Hall effect \cite{TypeIIQH}, magnetic breakdown and Klein Tunneling effect \cite{TypeIIMB}, Landau level collapse effect \cite{TypeIILL} have been predicted. ARPES measurements have been reported on MoTe$_2$ \cite{MoTeARPESKaminski, MoTeARPESZhou, MoTeARPESChen, MoTeARPESGrioni, MoTeARPESBaumberger}, WTe$_2$ \cite{WTeARPESBaumberger, WTeARPESKaminski1, WTeARPESRader, WTeARPESZhou, MoTeARPESKaminski2, WTeARPESSarma} and Mo$_x$W$_{1 - x}$Te$_2$ \cite{MoWTeARPES1, MoWTeARPES2}. Pump-probe ARPES measurements have discovered Weyl cone-Fermi arc connectivity, thus confirming the Weyl semimetal phase in the material \cite{MoWTeARPES2}. Furthermore, the spin texture on the surface states of WTe$_2$ \cite{WTeARPESSpin} and MoTe$_2$ \cite{MoTeARPESSpin} have also been discerned. Meanwhile, transport measurements have discovered the anisotropic Adler-Bell-Jackiw anomaly effect in WTe$_{1.98}$ crystals \cite{TypeIICA}. The discovery of MoTe$_2$ to be superconducting, whose T$_C$ can be enhanced by high pressure or doping with S \cite{TypeIISC1, TypeIISC2} may open a route towards the realization of an unconventional topological superconductor.

\begin{figure*}
\centering
\includegraphics[width=10cm]{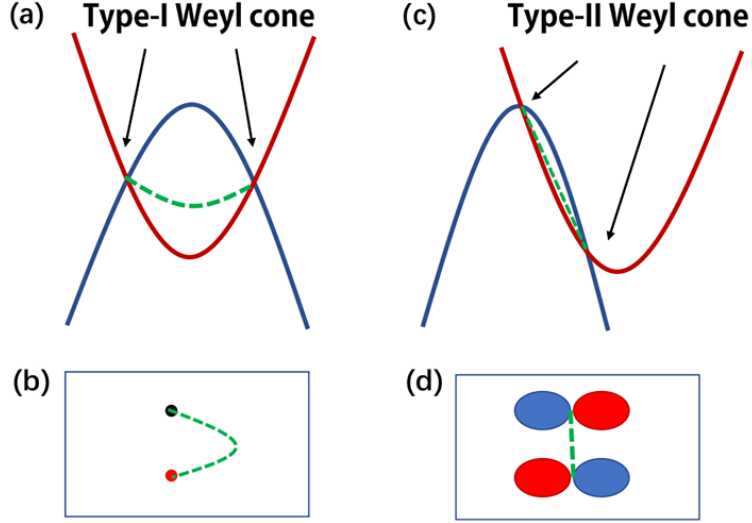}
\caption{\label{Fig1} \textbf{Type-I and type-II Weyl Semimetals}
(a)	A sketch of the E-k dispersion of a type-I Weyl semimetal which can be viewed as a negative direct bandgap semiconductor. The type-I Weyl nodes exist at the crossing points where the conduction band (red) and valence band (blue) dip into each other. A Fermi arc (green dotted line) connects the pair of Weyl nodes. (b) The Fermi surface (FS) on the surface of a type-I Weyl semimetal as depicted in (a). The Fermi level is located at the energy of the Weyl nodes. Bulk Weyl cones are projected onto the surface as two discrete points (red, blue) connected by a Fermi arc surface state (dotted green line). (c) A sketch of E-k dispersion of an inversion symmetry breaking type-II Weyl semimetal which can be viewed as a negative indirect bandgap semiconductor. Tilted type-II Weyl cones form at the intersection of conductance and valence bands. (d) FS on the surface of a type-II Weyl semimetal as depicted in (c). Inversion symmetry breaking constrains the minimum number of Weyl nodes to be four. Here, the Fermi level is assumed to cut at the energy of the lower Weyl node in (c). Projected bulk electron (red) and hole (blue) pockets coexist on the surface FS and touch at one discrete point, the type-II Weyl nodes. A Fermi arc (green dotted line) connects one pair of projected Weyl nodes.
}
\end{figure*}

From a material science perspective, the type-I Weyl semimetal can be treated as a direct negative band gap semiconductor while the type-II Weyl semimetal features an indirect negative gap as shown in Figure 1. As one might expect, many more type-II Weyl semimetals than type-I exist in real materials. The discovery the Weyl semimetal phase in TaAs- and WTe$_2$-class materials has attracted tremendous research effort in the search for new Weyl semimetals. Experiments have verified the existence of several new theoretically predicted Weyl semimetals including LaAlGe \cite{LAG}, TaIrTe$_4$ \cite{TITTheory, TITBorisenko, TITHasan}, Ta$_3$S$_2$ \cite{TaSHasan, TaSChen}, WP$_2$-class \cite{MoP, MoPBalicas},  Mn$_3$Sn-class \cite{MnSn1, MnSn2, MnSnNakatsuji}, and $\beta$-Ag$_2$Se \cite{AgSe}, which largely belong to the type-II Weyl semimetal category. Research into Weyl semimetals both in the context of new materials and new phenomena is, at present, an extremely dynamic field. This concise review article will focus on the TaAs-class and WTe$_2$-class of materials. As discussed above, while ARPES experiments were employed to identify the Weyl semimetal phase in real materials, electronic transport and optical measurements can, for the most part, only detect bulk band induced effects. As such, directly observing the physics that arises from the Fermi arcs requires a surface sensitive approach. Scanning tunneling microscopy (STM) combined with spectroscopy, which possesses extremely high spatial and energy resolution and probes the surface electronic structure of a crystal, is a natural choice. Here, we summarize the recent advances in STM-based quasiparticle interference (QPI) results on two prototypical type-I and type-II Weyl semimetals. This review article is organized as following: 1) the concept of Berry phase, a building block in the theory of all topological materials, is first introduced; 2) followed by the basics of STM and QPI; 3) and finally, concrete QPI results on both types of Weyl semimetals are presented.

\section{The concept of Berry phase, Weyl cone and Fermi arc}

Berry phases were initially introduced to describe the geometrical phase acquired by a particle’s wavefunction under an adiabatic variation of an applied external field. Later, research showed that the Chern number, an integer number that determines the number of quantum transport channels on the edge of a sample in a quantum Hall regime, is directly related to the Berry phase. Further investigations discovered that all symmetry protected topological phases in condensed matter can be described by the Chern number and Berry phase under various circumstances. We start from the definition of the Berry vector potential in a crystalline solid system.
\begin{equation}
\vec{A(k)} = i \left\langle k \left|\vec{\nabla _k} \right| k \right\rangle
\end{equation}
Where $|k \rangle$ is the Bloch wavefunction in k space. The integration of the Berry vector potential around arbitrary closed $c$ in $k-$space gives the Berry phase $\gamma$.
\begin{equation}
\gamma = \oint_c \vec{A(k)} \cdot d\vec{k}
\end{equation}

As we know from electrodynamics, the curl of a magnetic vector potential gives the strength of the magnetic field. Here, we apply the same concept to calculate the curl of the field strength of the Berry vector potential; the Berry curvature. Through Stokes theorem, we have:

\begin{equation}
\gamma = \oint_c \vec{A(k)} \cdot d\vec{k} = \int \int_S (\vec{\nabla_k} \times \vec{A(k)}) \cdot d\vec{S_k}
\end{equation}

Here, the surface $S$ in $k-$space is surrounded by loop $c$. If the integration is over a closed surface in $k-$space, for example, a Brillouin zone (BZ) in a two-dimensional material, or a closed surface inside of the BZ in a three-dimensional material, we obtain the Chern number $C$ which determines the number of edge states on the boundary.

\begin{equation}
C = \frac{1}{2\pi} \oint_S (\vec{\nabla_k} \times \vec{A(k)}) \cdot d\vec{S_k}
\end{equation}
We now describe a two level system with Hamiltonian $ H = \vec{d}(k_x,k_y,k_z) \cdot \vec{\sigma}$, where $d_{x,y,z}$ are functions of $(k_x,k_y,k_z)$ and $\sigma_{x,y,z}$ are the Pauli matrices.. Its eigenstate intrinsically possesses a nontrivial Chern number as the Berry curvature is calculated to be $\vec{\nabla_k} \times \vec{A(k)} = \frac{1}{2} \frac{\vec{d}}{d^3}$. From here, we can recognize this to be a Berry curvature monopole in $k-$space, and the Chern number to be the flux of the Berry curvature field on the integration plane.

While, historically, the Weyl semimetal was discovered after the topological insulator, topological crystalline insulator, and Dirac semimetal, the Weyl semimetal is perhaps the simplest symmetry protected topological phase in a mathematical sense. Unlike the required time reversal symmetry in TIs, mirror symmetry in TCIs, and rotational symmetry in Dirac semimetals, a Weyl semimetal only requires translational symmetry to preserved, a condition naturally fulfilled by any crystalline solid. In this sense, a Weyl semimetal may be the most robust symmetry protected topological phase. A simple model Hamiltonian which describes a Weyl semimetal containing only a single pair of Weyl nodes can be written
\begin{equation}
H(\vec{k}) = 2t_1sin(\frac{k_x}{2})\sigma_x + 2t_1sin(\frac{k_y}{2})\sigma_y + t_2(cosk_z - m)\sigma_z
\end{equation}

In the Hamiltonian, the two energy bands cross each other at the two $(0, 0, \pm arccos(m))$  points in momentum space.
In the vicinity of these two degeneracy points, we can expand the Hamiltonian in Equ.5 and get $h = \pm(v_xk_x\sigma_x +v_yk_y\sigma_y+ v_zk_z\sigma_z)$.
This is nothing but a Weyl fermion with a speed of $\vec{v}$ instead of light speed.

\begin{figure*}
\centering
\includegraphics[width=12cm]{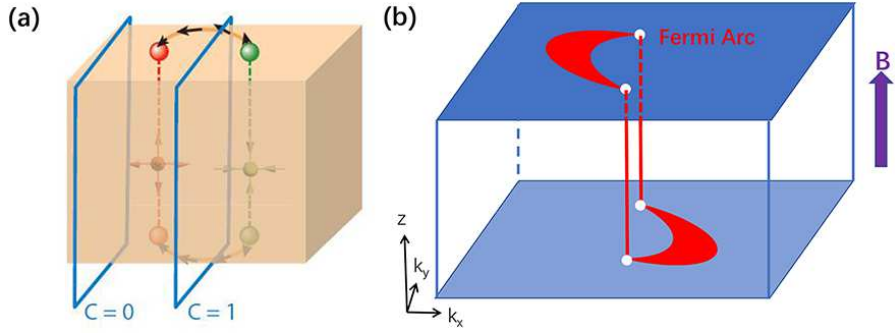}
\caption{\label{Fig1} \textbf{Demonstration of the Fermi arc and the topological sink effect}
(a) In momentum space, the $k_z$-dependent values of the Chern number generates the interrupted-shape surface state terminating at the projected position of the Weyl nodes—this is the Fermi arc surface state \cite{WeylRevYan}. (b) By placing a Weyl semimetal under a magnetic field perpendicular to the $z$ direction, a surface electron will travel through a Fermi arc surface state, reach the projected Weyl node, sink into the bulk state, move to the opposite surface, and finish the closed loop. This topological sink effect is the key phenomena associated with the unclosed shape of the Fermi arc.
}
\end{figure*}

The Hamiltonian in Eq. 5 can be written in the form $ H = \vec{d}(k_x,k_y,k_z) \cdot \vec{\sigma}$. It is thus topologically non-trivial as its wavefunction features non-trivial Chern numbers. Interestingly, if one chooses a $(k_x, k_y)$ plane located in the range of $-\arccos(m) < k_z < \arccos(m)$ as the integration plane of the Berry curvature, a Chern number of 1 is calculated, indicating the existence of one edge mode in this region (a surface state in three-dimensions), while integrating on the planes of either $-\arccos(m)>k_z$ or $k_z>\arccos(m)$ results in a Chern number of zero, describing the absence of an edge mode in this region. It is this $k_z-$dependent Chern number which generates the unique shape of the surface state as shown in Figure. 2. The surface state occurs only in the region between the pair of projected Weyl nodes on the surface and has an unclosed contour shape and is thus accordingly called a Fermi arc.

During the discovery of type-II Weyl Weyl semimetal, the researchers found that a Weyl cone can be described as $H = \sum k_i A_{i,j} \sigma_j$ ($i = x, y, z; j =0, x,y,z$) with energy, momentum-dispersion as $E_{\pm}(k) = \sum_i k_iA_{j,0}\pm \sqrt{\sum_i(\sum_j k_i A_{i,j})^2} = T(k) \pm U (k)$.
T(k) and U(k) can be considered as the kinetic and potential components of the dispersion.
$T(k) > U(k)$ titles the Weyl cone, thus leads to the type-II Weyl semimetal.
The generalized Weyl Hamiltonian is certainly beyond the original Weyl equation $H = \pm \vec{k}\cdot \vec{\sigma}$, but still has the form of $ H = \vec{d}(k_x,k_y,k_z) \cdot \vec{\sigma}$.
Therefore, both types of Weyl cones posses same Fermi arc type surface states. 

\section{Basics of Quasiparticle Interference Experiment and Theory}

Measuring the surface state on a crystalline solid requires a surface sensitive measurement. The principle behind STM lies in measuring the tunneling current between a sharp metal tip brought very close (< 1 nm) to a clean crystal surface. The relationship between the tunneling current and the applied voltage and distance between the tip and sample is as follows (under a one-dimensional barrier Bardeen tunneling model at zero temperature):
\begin{align}
&I = \frac{4\pi e}{\hbar}\int_0^{eV}\rho_T (\epsilon-eV) \rho_S (\epsilon) \left|M(\epsilon) \right|^2 d\epsilon\\
&\left|M(\epsilon) \right|^2 \propto exp(-2d \sqrt{\frac{2m}{\hbar}\Phi})	
\end{align}
In most cases, the first derivative (dI/dV) of the tunneling current is proportional to the local density of states of the sample $\rho_s$. Thus, a dI/dV(x,y) conductance map can directly measure the real space distribution of the local density of states. A surface state on a crystal can be treated as a special kind of two-dimensional electron gas which can be scattered at local point defects on the surface such as an atom vacancy or adatom. The incident wave vector ($\vec{k_i}$) and reflected wave vector ($\vec{k_r}$) can interfere at the site of these local point defects, leading to two-dimensional standing wave patterns. These patterns display different wave lengths at distinct energies, which is known as quasiparticle interference (QPI). A Fourier transform (FT) is applied to the real space dI/dV maps in order to gain insight in momentum space. A FT-dI/dV map plots all allowed surface scattering vectors ($\vec{Q}=\vec{k_r}-\vec{k_i}$). In combination with theoretical calculations, QPI measurements can provide rich information on the surface band structure both in occupied and unoccupied states, as well as the allowed and forbidden scattering channels of surface states. QPI measurements have been extensively employed in the investigation of surface states in many different topological materials \cite{QPI1Yazdani, QPI2Yazdani, QPI3Bode, QPI4Yazdani, QPI5Yazdani, QPI6Madhavan, QPI7Kimura, QPI8Yazdani, QPI9Bode, QPI10Stroscio, QPI11Wiebe, QPI12Komori, QPIDavic, QPI13Xue, QPI14Madhavan, QPI15Komori}. On the surface of a TI, QPI results demonstrate that the back scattering of surface carriers are forbidden at time reversal invariant local defects while they are allowed at magnetic defects, a signature manifestation of the unusual spin-momentum locking in a topological surface state. On the surface of TCIs, the unconventional orbital texture of the surface Dirac cone at different valleys has been successfully detected by QPI.

When acquiring QPI measurements on a crystal surface, a few technical issues need to be taken into account. On a surface, local defects will also induce a defect state which imposes an additional signal to that of the lattice. When this signal is included in the Fourier transform to Q-space, it can complicate the interpretation of the FT-dI/dV map. However, as QPI patterns measure surface standing waves which usually disperse with energy, defect states tend not to be dispersive and can thus be distinguished from the surface states. Secondly, there are two manners in which to obtain QPI maps: a dI/dV grid, also referred to as constant current tunneling spectroscopy (CITS), and dI/dV maps. A CITS grid is obtained by measuring the energy dependent dI/dV spectra on each real space point over an entire image, in order to obtain a set of local density of state maps at different energies. A dI/dV map simultaneously measures a dI/dV signal at a single energy, along with a constant current STM image. By varying the set point voltage, a series of dI/dV maps can be individually obtained. Furthermore, as shown above, the tunneling current is the integration of the sample’s local density of states from the Fermi level to the set point voltage. Even under a small voltage, the variation in the spatial local density of states can induce large changes in the tunneling current. To maintain a constant-current mode, the tip of the STM will withdraw from or approach the sample accordingly. This change in the tip-to-sample distance will affect the strength of the dI/dV signal and is referred to as a set-point effect. A typical set-point effect manifests as an additional duplicate and weakly dispersed pattern in the FT-dI/dV maps \cite{Setpoint}. A dI/dV grid measurement set at large voltages tends to result in a relatively smaller set-point effect.

Theoretical calculations of the QPI of a particular material is based on a density functional theory (DFT) simulated surface electronic band on a slab geometry. An autocorrelation of the surface Fermi surface results in a joint density of state (JDOS). By prohibiting spin-flip scattering, a spin-dependent scattering probability (SSP) is generated and can be applied to interpret the experimental QPI results \cite{TaAsQPIBernevig}. A more comprehensive method of simulating the QPI signal is the so-called T-matrix model, which is believed to capture the interference effects on the surface electronic state \cite{TaAsQPIFritz}. In the T-matrix method, the governing equations for the QPI can be written as \cite{TaAsQPIChang}
\begin{align}
&QPI(\vec{Q}, \omega) = \frac{i}{2\pi}\int\frac{d^2k}{(2\pi)^2}[B(\vec{Q},\omega)-B^*(\vec{Q},\omega)]\\
&B(\vec{Q}, \omega) = \textbf{Tr}[G(\vec{k},\omega)T(\vec{k},\omega)B(\vec{k}+\vec{Q},\omega)]\\
&T(\vec{k},\omega) = [1-V_i\int\frac{d^2k}{(2\pi)^2}G(\vec{k},\omega)]^{-1}V_{i}
\end{align}
where $\vec{k}$ and $\vec{Q}$ are the wave vector and scattering vector respectively, $G(\vec{k},\omega)$ is the surface Green function, $T(\vec{k},\omega)$ is the T-matrix, and $V_i$ is the defect induced potential.

\section{Quasiparticle interference on type-I Weyl semimetal}
The TaAs-class of materials which includes TaAs, TaP, NbAs, and NbP all share the same face-centered tetragonal crystal structure with space group $I4_1md$ which lacks inversion symmetry, as well as similar bulk and surface electronic bands. In the bulk electronic structure with no SOC, the conduction and valence bands of these materials intersect on four closed loops in the first bulk BZ. The inclusion of SOC gaps out this these closed loops, but the bands still intersect at twelve pairs of discrete points, each of which is a pair of Weyl nodes with opposite chirality. On the (001) surface of TaAs, the 24 Weyl nodes project onto 16 points. Eight of these projected Weyl nodes (W1) near the surface BZ boundary ($\bar{X}$ and $\bar{Y}$) carry a projected chiral charge of $\pm$1. However, as these W1 Weyl nodes are located very near to each other in k-space, the Fermi arcs associated with these Weyl nodes is difficult to distinguish from the trivial surface state. The other eight projected Weyl nodes near the $\bar{\Gamma}$ point have a projected chiral charge of $\pm$2 and are referred to as W2. As demonstrated in Figure 3c, the surface Fermi state of NbP(001) can be grouped into three types of pockets according to their differing contour shapes: a bowtie-shaped contour near $\bar{X}$ and an elliptical contour near $\bar{Y}$ which surround W1 Weyl nodes and are dominated by topological trivial states, as well as a tadpole-shaped contour pointing to the $\bar{\Gamma}$ point. The head of this tadpole-shaped contour connects one pair of projected bulk W2 Weyl nodes with projected chiral charges of $\pm$2, and is thus identified as the Fermi arc.

\begin{figure*}
\centering
\includegraphics[width=10cm]{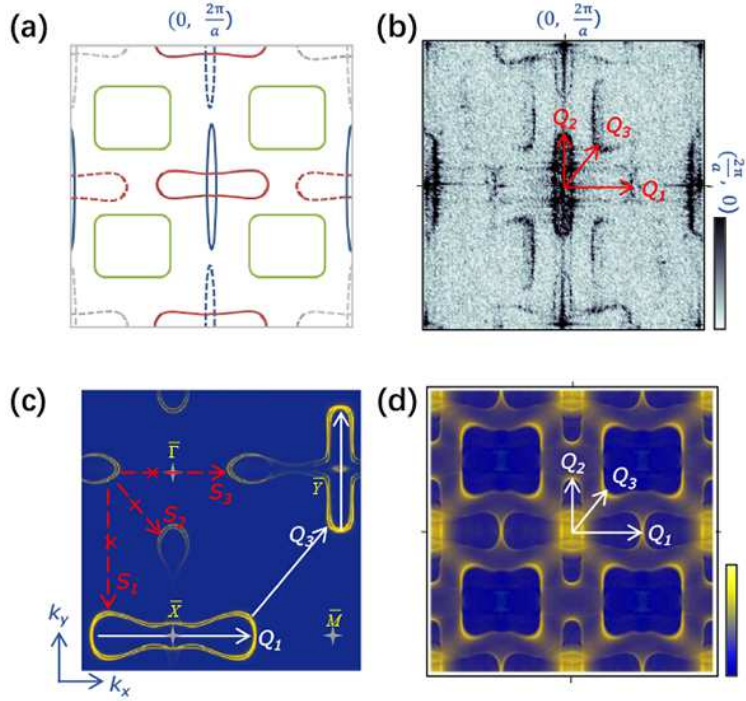}
\caption{\label{Fig3} \textbf{Scattering channels on NbP(001) surface}
(a) Sketch of the QPI pattern on NbP(001). The QPI features can be categorized into three groups, namely the bow-tie shaped (red), ellipse (blue) and rounded rectangle contours. (b) Experimental QPI pattern measured near the Fermi level. The three dominant scattering vectors are marked. (c) First principle simulation of the surface Fermi surface of NbP(001). The intra-bow-tie, intra-ellipse, and inter contour scattering vectors are indicated, corresponding to the three scattering vectors in (b). (d) Theoretically calculated QPI pattern which corroborates the measurement \cite{TaAsQPIZheng1}
}
\end{figure*}

As the TaAs-class of materials lack inversion symmetry, the P/As-surface is distinct from the Ta/Nb surface. In STM, prior to QPI measurements, dI/dV spectra taken on the surface are compared to the theoretical calculated LDOS in order to determine the surface orientation. To date, all reported STM experiments have been performed on the anion-terminated (001) surface \cite{TaAsQPIZheng1,TaAsQPIYazdani1,TaAsQPIBeidenkopf,TaAsQPIBode,TaAsQPIYazdani2,TaAsQPIZheng2}. Zheng \textit{et. al} reports QPI results on the NbP(001) surface \cite{TaAsQPIZheng1}. Voltage dependent dI/dV maps show surface standing waves increasing in wavelength, demonstrating the hole-like nature of these states. The main features in the FT-dI/dV map taken near Fermi level (V=-10mV) in Figure 3b can be divided into three groups: 1) a bow-tie shaped contour at the center of the image the end of which is marked by the vector $Q_1$; 2) an elliptical contour in the center of the image and perpendicular to the bow-tie. The vector $Q_2$ points to its end; 3) a rectangular feature with curved edges in each quadrant of the image. The corner is indicated by $Q_3$. The features at the Bragg spots can be viewed as replicas of the bow-tie and elliptical features. A model calculation taking matrix element effects into account (Figure 3d) reproduces the features in the experimental QPI pattern well. Owing to this agreement with these theoretical results of the QPI contours (Figure 3b) and the surface Fermi surface pocket (Figure 3c), the three dominant scattering vectors can be unambiguously identified. $Q_1$ and $Q_2$ arise from the intra-contour scattering, while $Q_3$ is the inter-contour scattering. The tadpole to  bowtie ($S_1$), and tadpole to a perpendicular tadpole ($S_2$) scattering are absent from experiment, which is attributed to the different orbital character of the contours.

\begin{figure*}
\centering
\includegraphics[width=14cm]{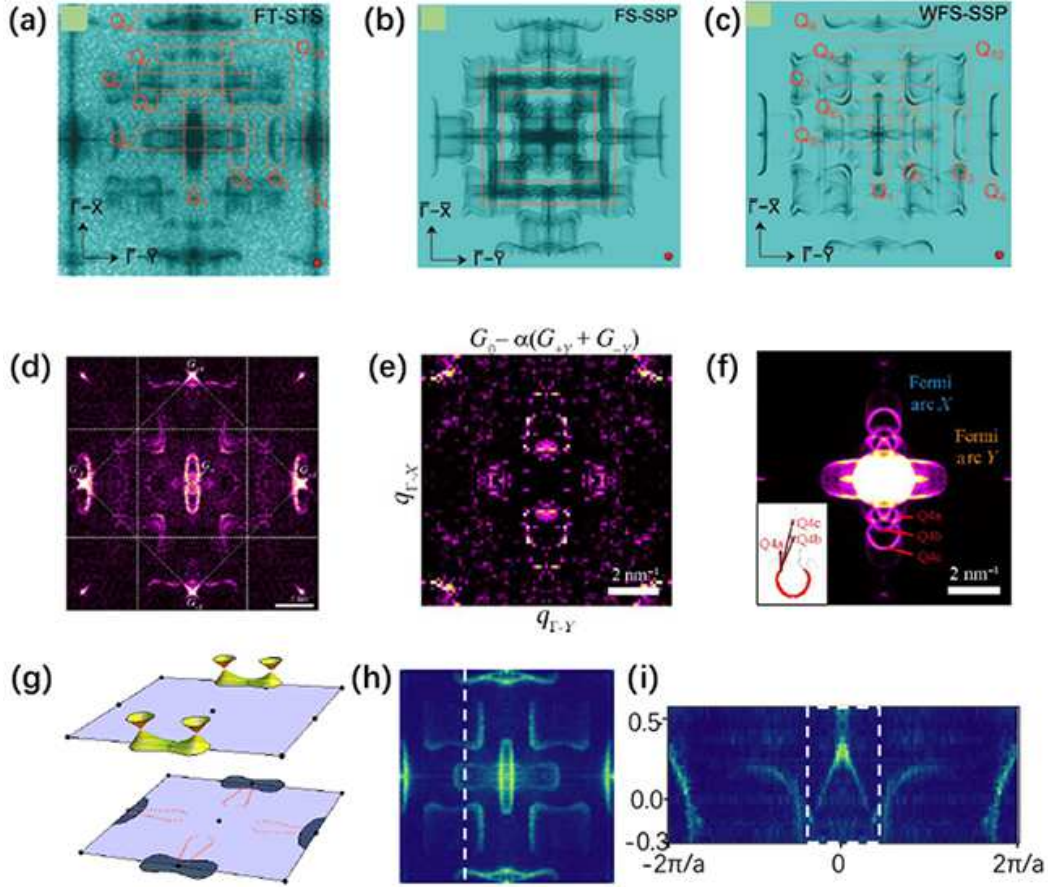}
\caption{\label{Fig4}
\textbf{Summary of recent QPI results on TaAs and NbP type-I Weyl semimetal surfaces}
(a) an experimentally acquired QPI pattern on TaAs(001) surface.
(b) a theoretical calculation of the QPI pattern with taking into account of the entire surface unite contribution.
(c) same as (b) but only considering the topmost As layer.
Obviously, (c) fits to (a) better than (b) \cite{TaAsQPIYazdani1}.
(d) an experimentally acquired QPI pattern on TaAs(001) surface from anther report \cite{TaAsQPIBeidenkopf}.
(e) a subtraction of the QPI contour at Bragg points from the central point.
It is believed to demonstrate the interference from Fermi arc surface state.
(f) a theoretical calculation of the QPI pattern of (e).
(g) a sketch of the two-dimensional Dirac cones (upper panel) and the Fermi surface (lower panel) in the first surface BZ of NbP(001) \cite{TaAsQPIZheng2}.
(h) an experimentally acquired QPI pattern on NbP(001) surface.
(i) an energy-scattering vector (E-Q) dispersion cut along the dotted line in (h).
A Dirac cone type feature is revealed.
}
\end{figure*}

Inoue \textit{et al.} report QPI results on a TaAs(001)-As surface \cite{TaAsQPIYazdani1}, identifying the scattering vector ($Q_2$ in Figure 4a) linking the tadpole and elliptical contours. They find the bulk state in the vicinity of the Weyl cones arises mainly from the Ta-\textit{d} orbital and the surface state mainly from the As-\textit{p} orbital. Simulated QPI patterns were calculated based on two different geometries, namely on a single surface unit consisting of four Ta-As bilayers, and on only the topmost As-layer. A comparison between the measurement and simulation indicates that only the top As-layer is observed in QPI patterns. This is interpreted as the electron from a Fermi arc surface state (on an As site) moving to the Ta site entering the bulk Weyl cone, leaving the surface and sinking into the bulk. This result serves as an important signature of the unique topological sink effect of a Fermi arc state.

Batabyal \textit{et al.} also measured the TaAs(001)-As surface QPI \cite{TaAsQPIBeidenkopf} and found similar QPI patterns as Inoue \textit{et al.} as shown in Figure 4d. Through a weighted subtraction of the bowtie and elliptical features at the Bragg points from the center region, they discovered that the residual QPI features are induced by scattering between the Fermi arc to the nearby bulk state. As these features do not repeat themselves at Bragg points, the authors interpret it as the electronic state which induces these scatterings possessing only long wave-length standing waves in real space rather than atomic scale corrugations. Thus, they postulate that the Fermi arc surface state is weakly bonded to the lattice in contrast to a topologically nontrivial surface state.

Zheng \textit{et al.} have theoretically predicted two-dimensional Dirac cone states on the surface of NbP(001) which are protected by the $\bar{X}-\bar{\Gamma}-\bar{X}$ mirror symmetry \cite{TaAsQPIZheng2}. A pair of surface Dirac cone states are located near the $\bar{\Gamma}$ point in the surface BZ, where the Dirac nodes are about 300meV above the Fermi level. As the energy moves towards the Fermi level, the two Dirac cones expand in size and eventually merge and evolve into the bow-tie shaped contour at the Fermi level. The Dirac nodes exist in the unoccupied states, and thus cannot be accessed by conventional ARPES measurements. Energy resolved QPI measurements were able to detect the surface Dirac cone. As shown in Figure 4i, an energy-scattering vector ($E-Q$) dispersion measurement on the NbP(001) surface reveals a $\Lambda$-shaped feature in the center, proving the existence of the predicted mirror protected Dirac cone surface state.

\section{Quasiparticle interference on type-II Weyl semimetal}

It was only recently that type-II Weyl semimetals were predicted. To date, the most intensively studied type-II Weyl semimetals are those in the WTe$_2$-class of transition metal dichalcogenides which includes WTe$_2$, MoTe$_2$, and their alloy Mo$_x$W$_{1-x}$Te$_2$. Both WTe$_2$ and MoTe$_2$ feature type-II Weyl cones in their bulk bands when they crystallize in the T$_d$ structure polymorphy. Although the number of Weyl nodes (4 or 8) are sensitive to tiny variations in the lattice parameters used in first principle calculations, all simulations agree that the Weyl cones are located above the Fermi level. Therefore, it is beyond the capability of conventional ARPES measurements to detect the Weyl cone and the Fermi arc state at the energy of the Weyl node. However, QPI measurements do not suffer such constraints \cite{MoTeQPIZheng, MoTeQPIChen, MoTeQPIWu, MoTeQPITakagi}. Figure 5a demonstrates the entire Fermi surface of Mo$_{x}$W$_{1-x}$Te$_2$ (001) surface. The type-II nature of the Weyl cones results in the coexistence of projected bulk and surface states. The bulk state consists of one dog-bone-shaped hole pocket in the center of the BZ and two elliptical electron pockets near the left and right edges. The projected Weyl nodes (black and white dots in Figure 5a) are located at the boundary between the hole and electron pockets. The two bright yellow semi-circular contours in Figure 5a are surface states, where the middle segments are the Fermi arc surface states. Theoretical analysis has revealed that the Fermi arc surface states are derived from the Mo-$d$ orbitals. In other words, the Mo$_{x}$W$_{1-x}$Te$_2$ alloy may enhance the surface electron scattering at metallic sites, consequently enhancing the Fermi arc interference signal. The theoretical QPI calculated by considering both the projected bulk and surface states and generates a pattern consisting of seven pockets (Figure 5b), while a surface state only QPI consists of only three pockets (Figure 5d). The experimental QPI is consistent with the result from Figure 5d, indicating that the projected bulk states do not factor into the QPI. This is interpreted as a surface electron which is sitting on a Fermi arc moving to the projected Weyl cone pocket and sinking into the bulk, and thus does not contribute to the surface standing waves, the signature of the topological sink effect in a type-II Weyl semimetal. Due to the tilted nature of the type-II Weyl cone, the area of the projected Weyl bulk states in a type-II WSM is much larger than in type-I, which exists at a single discrete point. Therefore, the topological sink effect is in principle much more pronounced on a type-II WSM. Both measured QPI patterns on WTe$_2$ and MoTe$_2$ in Figure 5 manifest themselves as three pocket structures as observed in Mo$_{x}$W$_{1-x}$Te$_2$, indicating the robustness of the topological sink effect in this class of materials.

\begin{figure*}
\centering
\includegraphics[width=10cm]{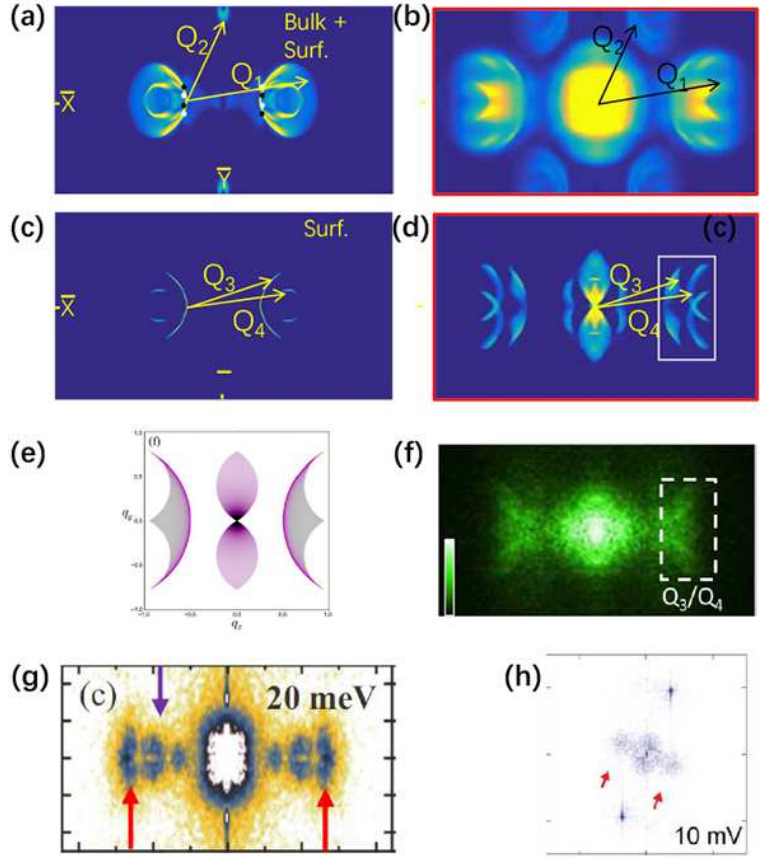}
\caption{\label{Fig1} \textbf{QPI results on WTe$_2$-class of type-II Weyl semimetal surfaces.}
(a) Calculated ``complete" Fermi surface of a Mo$_{x}$W$_{1-x}$Te$_2$(001) surface. It contains both the bulk and surface state. (b) QPI pattern derived from (a), presenting only the intra-BZ scattering. (c) Similar to (a), but with only the surface states taken into consideration. The Fermi arc is the central segment of the semicircular contours. (d) QPI pattern derived from (c) \cite{MoTeQPIZheng}. (e) The theoretically simplest Fermi arc QPI pattern \cite{TaAsQPIBernevig}. (f) Experimental QPI data on Mo$_{x}$W$_{1-x}$Te$_2$(001), corresponding to (d). (g) Experimental QPI data on a WTe$_2$(001) surface \cite{MoTeQPIWu}. (h) Experimental QPI data on a MoTe$_2$(001) surface \cite{MoTeQPIChen}
}
\end{figure*}

Furthermore, simulations considering only the surface state (Figure 5c) reveals that on the (001) surface of Mo$_{x}$W$_{1-x}$Te$_2$, an ideal WSM is approximately realized, which would lead to the simplest Fermi arc QPI (Figure 5d). The theoretical simulation in Figure 5d is based on a simple analytical model without considering material specific parameters and shows only a single pair of Fermi arcs. The QPI features a butterfly shape with three QPI pockets, corroborating the measurements on Mo$_{x}$W$_{1-x}$Te$_2$. It thus renders this class of materials as an ideal real material for the investigation of Fermi arc interference patterns.

\section{Perspective and Outlook}

In the research of surface electronic states, ARPES possesses a unique position due to its distinct capability for directly detecting the dispersion of surface bands. However, conventional ARPES is only capable of probing the occupied states and is furthermore incompatible with magnetic field. On the other hand, scanning tunneling microscopy QPI measurements provide only indirect information on the surface band structure but are not restricted to the occupied states and are well suited for magnetic field dependent measurements. Weyl semimetals can be divided into two categories depending on whether they are inversion or time reversal symmetry breaking. The aforementioned TaAs and WTe$_2$ classes of materials both belong wo the inversion breaking Weyl semimetal. Recently, a time-reversal symmetry breaking Weyl semimetal phase was predicted in ferromagnetic Heusler half metals \cite{Heusler1, Heusler2}. Interestingly, magnetization of the samples along distinct crystalline directions generate different band topologies and consequently distinct surface Fermi arc states. This exotic topological response to external field is unfortunately not accessible by ARPES. Due to the non-vanishing bulk electronic state, the electronic transport measurements may also fail to track such surface effects due to a lack of surface sensitivity. QPI measurements may then be the most suitable approach for probing such phenomenon. In fact, most phenomena arising from topological surface electronic states on time-reversal symmetry breaking Weyl semimetals are possibly most readily discerned and discovered via this strong experimental technique of QPI measurement.

\section{Acknowledgments}
We thank S. Zhang, S. Huang and S. Xu for the helpful discussions.
H.Z. acknowledges the financial support from National Natural Science Foundation of China (Grant Nos. 11674226,  11790313) and the National Key Research and Development Program of China (2016YFA0300403).
M.Z.H. acknowledges the Gordon and Betty Moore Foundations EPiQS Initiative through Grant GBMF4547 and U.S. National Science Foundation (NSF) Grant No. NSFDMR-1006492.

\end{document}